\documentclass[preprint2, 10pt]{aastex}

\newcommand{\Jpb}{\mbox Jy~beam$^{-1}$}
\newcommand{\uJpb}{\mbox{$\mu$Jy~beam$^{-1}$}}

\newcommand{\sn}[1]{\mbox{s$^{#1}$}}

\newcommand{\uv}{\mbox{\em u-v}}
\newcommand{\Ra}[4]{\mbox{${#1}^{\rm h} \; {#2}^{\rm m} \; {#3}\fs{#4} $}}
\newcommand{\dec}[4]{\mbox{${#1}\arcdeg \; {#2}\arcmin \; {#3}\farcs{#4} $}}

\begin{document}

\title{The Crab Nebula's Moving Wisps in Radio}
\author{M. F. Bietenholz}
\affil{Department of Physics and Astronomy, York University, 
North York, M3J~1P3, Ontario, Canada}

\author{D. A. Frail}
\affil{National Radio Astronomy Observatory, Socorro, New Mexico, 87801, USA}

\author{J. J. Hester}
\affil{Department of Physics and Astronomy, Arizona State University, Tempe, Arizona, 85287, USA}

\shorttitle{Crab Wisps in Radio}
\shortauthors{Bietenholz, Frail \& Hester}
\slugcomment{submitted to Ap.\ J.}


\vfill\eject
\begin{abstract}

We present three high resolution radio images of the Crab nebula,
taken in 1998.6, 1998.8 and 2000.1 with the VLA. These are the best
radio images of the Crab to date. We show that, near the pulsar, there
are significant changes between our three observing epochs.  These
changes have an elliptical geometry very similar to that of the
optical wisps.  One radio wisp in particular can be unambiguously
identified between two of our observing epochs, and moves outward with
an apparent velocity of $\sim 0.24 c$. The similarity in both
morphology and behavior of the present radio wisps to the optical
wisps suggests that they are associated.  This implies that the radio
wisps, like the optical ones, are likely manifestations of the shock
in the Crab pulsar's wind.  This suggests that the radio emitting
electrons are accelerated in the same region as the ones responsible
for the optical to X-ray emission, contrary to most current models.
\end{abstract}


\section{Introduction}

The \objectname[]{Crab Nebula} is one of the most important objects in
astrophysics.  It is a beautiful example of a pulsar powered nebula,
and is also the most accessible pulsar nebula.  Each generation of
observations has revealed new intricacies to its structure.  Recent
results have highlighted the remarkable nature of the region
surrounding the Crab Nebula's pulsar, where the pulsar wind is
injected into the nebula.  A remarkable image of this region from the
Chandra X-ray observatory (Weisskopf et~al.\ 2000) shows a complex
geometry consisting of a tilted toroidal structure with a jet along
the symmetry axis of the torus. A recent sequence of images from the
Hubble Space Telescope (HST) by Hester et al.\ (1996) revealed
astonishing details in this region.  In particular, there is a series
of elliptical ripples in this region, which are usually called
``wisps''.

The wisp region has long been known to be rapidly variable (Lampland
1921; Scargle 1969), but only recently has the nature of the
variability been conclusively established: Hester et~al.\ (1996; see
also Hester 1998) and Tanvir, Thomson, \& Tsikarishvili (1997) have
demonstrated that the wisps are variable on time-scales as short as a
few days and that they are moving outwards with apparent speeds of up to
$0.7 c$.  Early Chandra results indicate that the X-ray structure
is also variable, on timescales of $\sim 10$~days.

At the center of the Crab Nebula, and of the wisp region, the
rapidly spinning Crab pulsar is slowing down, and thus losing energy
at the rate of $5 \times 10^{38}$~erg~\sn{-1}.  This bulk of this
energy emerges from the pulsar as a highly collimated wind of
relativistic particles and magnetic field.  This outflow is randomized
at a shock, where it comes into equilibrium with its surroundings
(Rees \& Gunn 1974) and then flows into the body of the nebula, where
it gives rise to the synchrotron emission we observe from the body of
the nebula.  However, the process by which this transfer of energy
happens is not yet well understood (see e.g.\ Begelman 1999; Arons
1998; Hester 1998; Chedia et~al.\ 1997).

Since almost all the pulsar's spindown energy is accounted for in the
body of the nebula, we know that the transfer of this energy is quite
efficient.  Only a very small fraction of the energy seems to be
radiated directly from the region where the wind is randomized: the
wisps are now thought to be associated with the shock which terminates
the highly collimated outflow from the pulsar and randomizes it before
it enters the body of the nebula.  These features are our only
observational window into the process by which the pulsar outflow is
converted into the relativistic gas and magnetic field which fill the
body of the nebula.

A feature with an arc-like geometry similar to that of the optical
wisps was observed in the radio spectral index by Bietenholz \&
Kronberg (1992), who also observed changes in the radio emission from
the center of the nebula over a period of five years. Our goal with
these observations is to investigate the nature of the radio
synchrotron emission from the wisp region, and to compare it to the
synchrotron emission from the optical wisps.  In particular, we will
investigate whether the radio wisps are also rapidly variable, since
the extant observations of them show only that they are variable on a
5~year timescale.

\section {Observations and Data Reduction \label{sdatredux}}

We observed the Crab Nebula using the NRAO VLA\footnote{The NRAO Very
Large Array is a facility of the National Science Foundation operated
under cooperative agreement by Associated Universities, Inc.}  in the
5~GHz band.  Table~1 shows details of our three observing runs, all of
which were done using the VLA in the B configuration.  Due to
scheduling reasons, our observing run in February 2000 was split into
two segments, which we combined, and we refer to this combined data
set as our 11~February 2000 epoch.  At each of these sessions, we
observed at a spaced pair of frequencies (within the 5~GHz band) in
order to increase our \uv~coverage, since good \uv~coverage is
critical when imaging an object as extended as the Crab.  In addition,
we chose different pairs of frequencies for the 10~Feb.\ and the
12~Feb.\ 2000 observing runs, since these two runs had roughly the
same range in hour angle.

Our primary goal was to investigate any rapid variations in the
nebular structure.  Since a speed of $c$\/ represents a proper motion
of only $\sim 3\arcsec$ per month, we do not expect the large scale structure
of the nebula to change very much on short timescales.  We do know
that the radio synchrotron nebula is expanding at 0.13\%~year$^{-1}$
(Bietenholz et~al.\ 1991).  The effect of this expansion is both
predictable, since the rate is known, and is also quite small over the
timespan of our observations: the most rapid motion due to the overall
expansion rate is $< 0\farcs3$~yr$^{-1}$.

The fact that the large scale structure of the nebula is not expected
to change over our timespan is crucial to our strategy: an
interferometer like the VLA measures the spatial Fourier transform of
the sky brightness, and is sensitive only to a range of spatial
frequencies determined by the length of the interferometer baselines
and the observing frequency.  For example, in the B configuration at
5~GHz, the VLA is sensitive only to structure on spatial scales between
$\sim 1\farcs3$ and $\sim 50\arcsec$.  The Crab Nebula, however,
exhibits structure at spatial scales from $\sim 6\arcmin$ down to $<
1\arcsec$. Typically in such cases, one would obtain the larger scale
structure from VLA observations in more compact array configurations.
However, since VLA configuration changes occur only once every four
months, that approach is not compatible with our goal of obtaining
time-resolved observations.

Our solution to this problem is to use maximum entropy deconvolution
(AIPS task VTESS; see Cornwell \& Evans 1985, Cornwell 1988), which
allows us to recover the large scale structure by supplying a low
resolution support (default) in the deconvolution process.  The
deconvolved image is then biased to be as close to the support as is
allowed by the data.  We chose this support to be the same for all
three of our epochs\footnote{We did scale the support according to the
known expansion of 0.13\%~year$^{-1}$ but the effect of this is
minimal.} which serves to make any differences between the epochs be
only such as are demanded by the data.

As our support, we used an image of the Crab made using data taken in
1987, at 5~GHz, using the B, C and D configurations of the VLA
(Bietenholz \& Kronberg 1990; 1991).  We convolved this image with a
round Gaussian of full-width at half-maximum (FWHM) $20\arcsec$, and
scaled it to account for the general expansion of the radio nebula
since 1987.  Because we have only B~array data at the current epochs
of interest, the structure in our images on the largest spatial
scales, i.e.\ those $> 1\arcmin$, is derived predominately from the
support image, and we thus have very little information on any changes
that might occur at these spatial scales.  As we have argued above,
however, there is little reason to expect rapid variations on these
scales.  On spatial scales smaller than $\sim 30\arcsec$ on the other
hand, our \uv~coverage is excellent, and we believe that our images
reliably indicate the structure of the nebula, and in particular,
changes therein from one epoch to another.

\section{Results}

\begin{figure}
\plotone{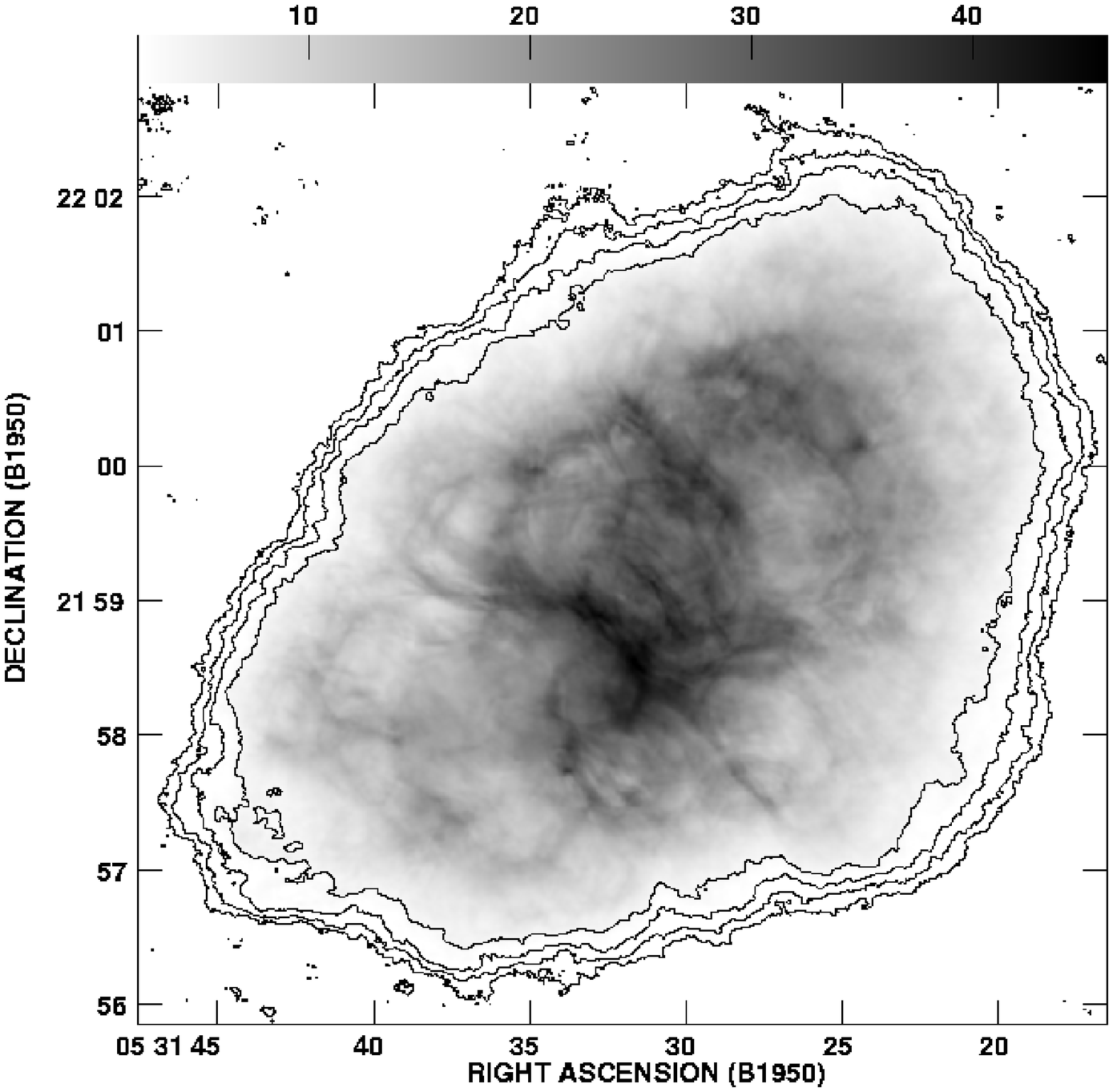}
\figcaption{The image of the Crab at 5~GHz on 2000 February 11 (from
data taken on Feb.\ 10 and 12), after primary beam correction. The
FWHM size of the restoring beam was 1\farcs4.  The peak flux density is
46.6~m\Jpb, and the exterior rms was 76~\uJpb\/ before primary beam
correction.  The contours are drawn at 0.75, 2, 4 and 8\% of the peak,
and the greyscale is in m\Jpb.  Maximum entropy deconvolution was used
with a support to recover the low spatial frequency structure (see
text, \S~\ref{sdatredux}) for details). \label{febmap}}
\end{figure}

An image of the Crab Nebula on 11~February 2000 is shown in
Figure~\ref{febmap}.  The restoring beam size was 1\farcs4 (this
restoring beam size was conservatively chosen as a common size for all
three epochs).  The background rms level was 76~\uJpb\ before applying
the primary beam correction, and the peak flux was 46.6~m\Jpb.  This
is the highest dynamic range and resolution radio image yet produced
of the Crab Nebula. We note that it is superior to the three and
four-configuration VLA images of Bietenholz \& Kronberg (1990; 1991)
--- on which it is partly based through the use of those images as a
support --- because of our presently increased \uv~coverage at the
higher spatial frequencies.

\begin{figure}
\plotone{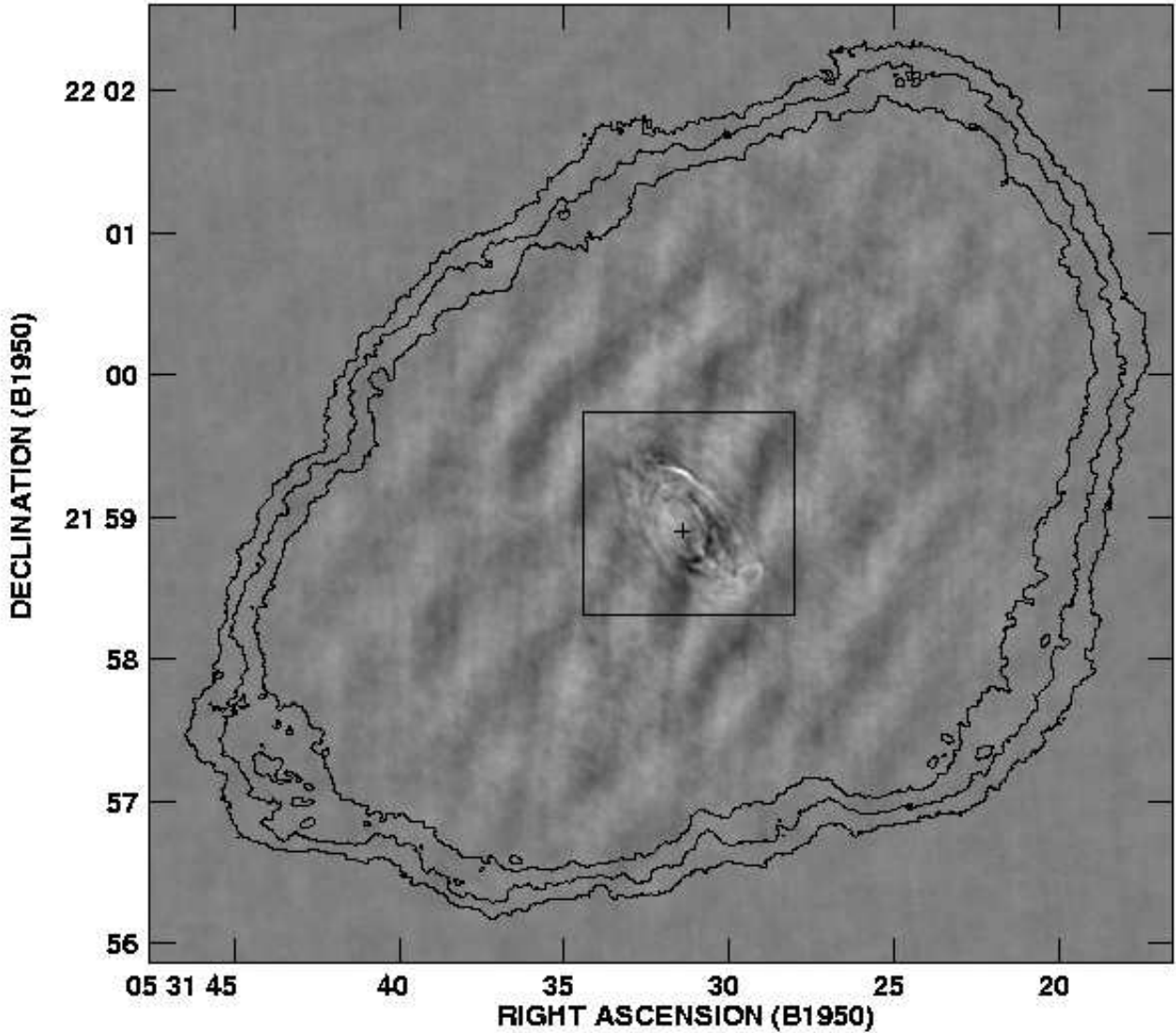} \figcaption{The differences between our 1998
October~13 and 1998 August~9 images in greyscale, with the greyscale
ranging from +5 to $-5$~m\Jpb.  The FWHM size of the restoring beam
was 1\farcs4.  For reference, we also show the the 2, 5, and 10\%
contours from the 2000 February~11 image, and indicate the pulsar
position by a cross.  The box indicates the region of the high-pass
filtered images shown below in Fig.~\ref{riplmaps}
\label{diffmap}}
\end{figure}

The images taken 1998 August and October were very similar, and so we
do not reproduce them in their entirety. They had background rms
levels of 100 and 82~\uJpb\ respectively.  Instead, we present a {\em
difference}\/ image: Figure~\ref{diffmap} shows the difference between
the images of October and August 1998.

\begin{deluxetable}{r@{ }l@{ }l c c c}
\tablewidth{0pt}
\tablecaption{Observing Run Details}
\label{obstab}
\tablehead{
\multicolumn{3}{c}{Date} & \colhead{Length} & \colhead{Frequencies} & \colhead{Midpoint} \\
  & & &
\colhead{(hrs)}  & \colhead{(MHz)} & \colhead{(days since 1998.0)}
}
\startdata
1998 & Aug & \phn9 & \phn 9.6 & 4615, 4885 & 221.7 \\
1998 & Oct & 13    &    10.1 & 4615, 4885 & 286.5 \\
2000 & Feb & 10    & \phn 3.7 & 4749, 4996 & 772.1 \\
2000 & Feb & 12    &\phn 5.7 & 4615, 4885 & 774.0 \\
(combined  &
       Feb & 2000)& \phn 9.4 &
4615, 4749, 4885, 4996                     &  773.1\tablenotemark{a} \\
\enddata
\tablenotetext{a}{The midpoint for the combined February 2000 runs was taken
to be the weighted mean of the midpoints of the two individual days.}
\end{deluxetable}

\begin{figure}
\epsscale{0.8}
\plotone{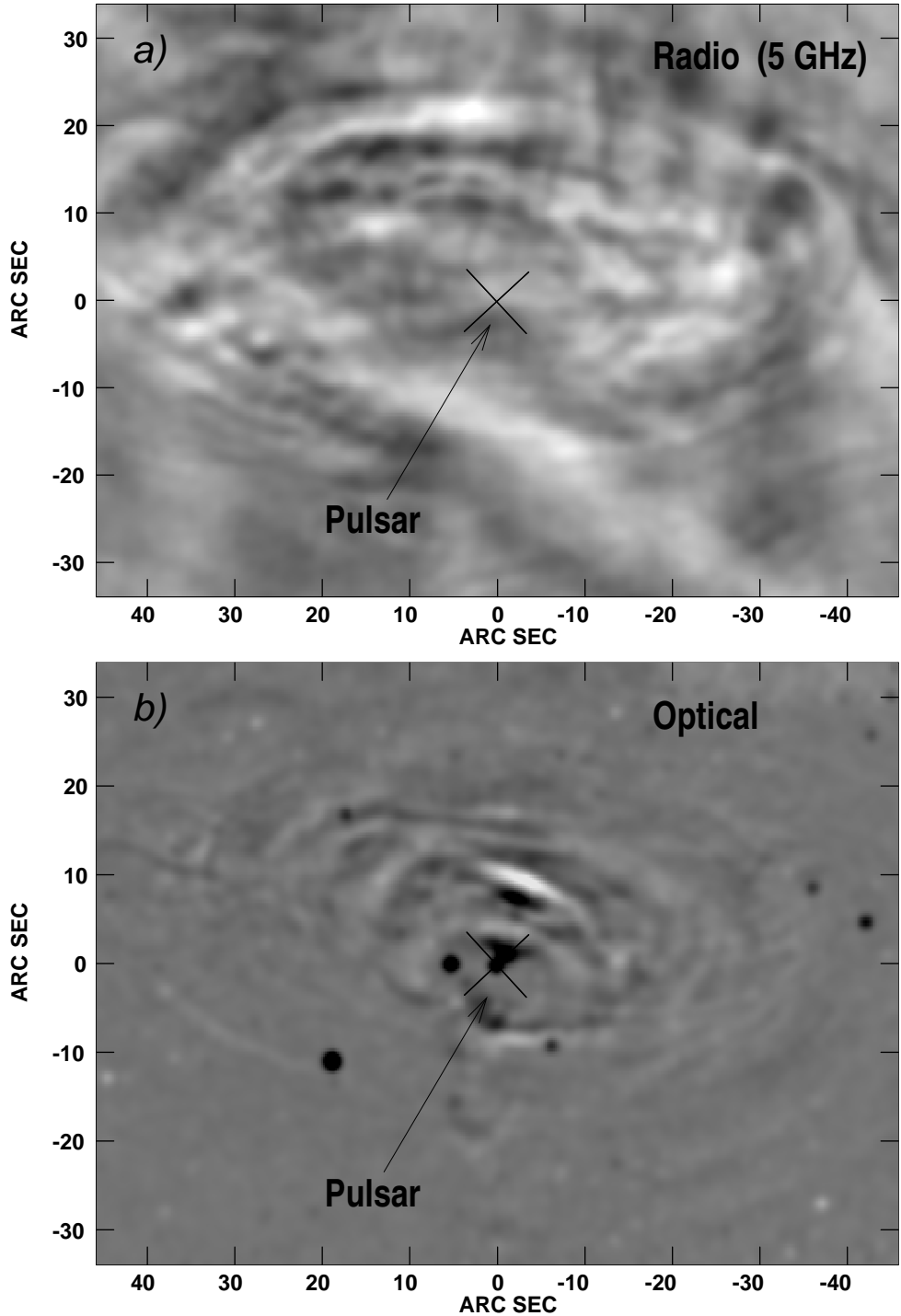}
\figcaption{The difference features in the optical and radio.  The
images have been rotated by $-47$\arcdeg.  The cross is drawn in the
N-S E-W direction, and shows location of the pulsar.  In $a)$ we show
the radio difference image between 2000 February~11 and 1998
October~13 (the greyscale ranges from +5 to $-5$~m\Jpb, and the
convolving beam size was 1\farcs4 FWHM).  In $b)$ we show the
difference between the HST images from 2000 October~23 and September
9, convolved also to 1\farcs4 FWHM resolution.
\label{optdiff}}
\end{figure}

On this difference image a prominent series of elliptical ripples are
visible near the center of the nebula.  The brightness of these
ripples is several times that of the background noise level.  Such
ripples are only visible near the pulsar position:
Figure~\ref{diffmap} shows that, in the rest of the nebula, any
difference features are of lower amplitude and larger spatial scale,
and likely due to noise, missing short-spacings, and errors in the
deconvolution.  The elliptical ripples show a remarkable similarity in
geometry with the elliptical wisps seen in the optical (Hester et al.\
1995; Hester et~al.\ 1996) and the torus visible in the X-ray
(Aschenbach \& Brinkman 1975; Weisskopf et~al.\ 2000).
Figure~\ref{optdiff} shows a region near the pulsar of the radio
difference image between 1998 October and 2000 February, and a
corresponding difference image between two HST images taken 1.5 months
apart and convolved to the same resolution (note that the images in
this figure have been rotated by $-47$\arcdeg).  In particular, the
implied inclination angle (assuming that the features are
intrinsically circular) is similar to that in both the optical and the
X-ray.  If the ripples were an artefact, such a coincidence in
geometry with the real features observed at other wavelengths would be
quite unlikely.  While moving features seem to exist over roughly the
same region in both the optical and radio images, the largest features
in the radio are further from the pulsar than the largest optical
features are.

\begin{figure}
\epsscale{0.45}
\plotone{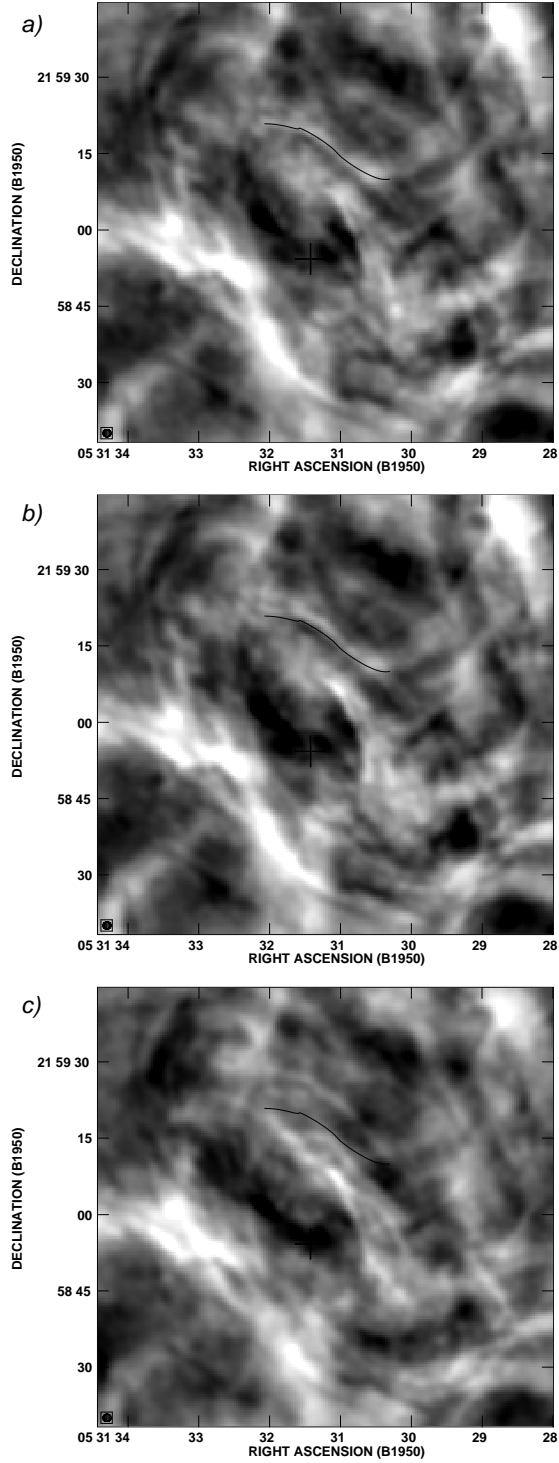} 
\figcaption{High-pass filtered images of the region
near the pulsar, showing the mobile wisps.  The images have been
high-pass filtered with a Gaussian of FWHM~25\arcsec.  The restoring
beam again has a FWHM size of 1\farcs4, shown at lower left.  The
position of wisp~$a$ on 1998 August~9 is indicated on all three
images to show its displacement.  In $a)$ we show the image from 1998
August~9, in $b)$ that from 1998 October~13, and in $c)$, that from
2000 February~11.
\label{riplmaps}}
\end{figure}

In order to make the motions from one epoch to the next better visible
in reproductions, we show in Figure~\ref{riplmaps} high-pass filtered
views of the center of each of our three images.  We used a Gaussian
high-pass filter with a FWHM of 25\arcsec, which serves to accentuate
the relatively faint wisp-like mobile features.  We will call the most
mobile feature wisp~$a$, and we indicate its position in 1998 August
by a black line on all three images in order to show the displacement
in the subsequent epochs.

These motions are real: they occur on spatial scales of a few
arc-seconds, exactly those to which we are sensitive and for which we
have excellent \uv~coverage.  In order to further ascertain that these
motions were not some form of artefact due to the deconvolution
process, we performed several tests.  Firstly, we deconvolved the
1998 August image using the October image as a support, instead of our
usual low-resolution image.  This further biases the images to be as
similar as is allowed by the data.  The ripples remained essentially
unchanged.  Secondly, expanding or contracting the default image by
2\% also left the ripples unchanged, but resulted in a poorer image.
Lastly, we self-calibrated the 1998 August \uv~data both in amplitude
and phase, using the October image as a model, and vice versa, and
then re-imaged.  Again, the ripples are essentially unchanged,
implying that they cannot be ascribed to any calibration anomalies.

\begin{figure}
\epsscale{1.10}
\plottwo{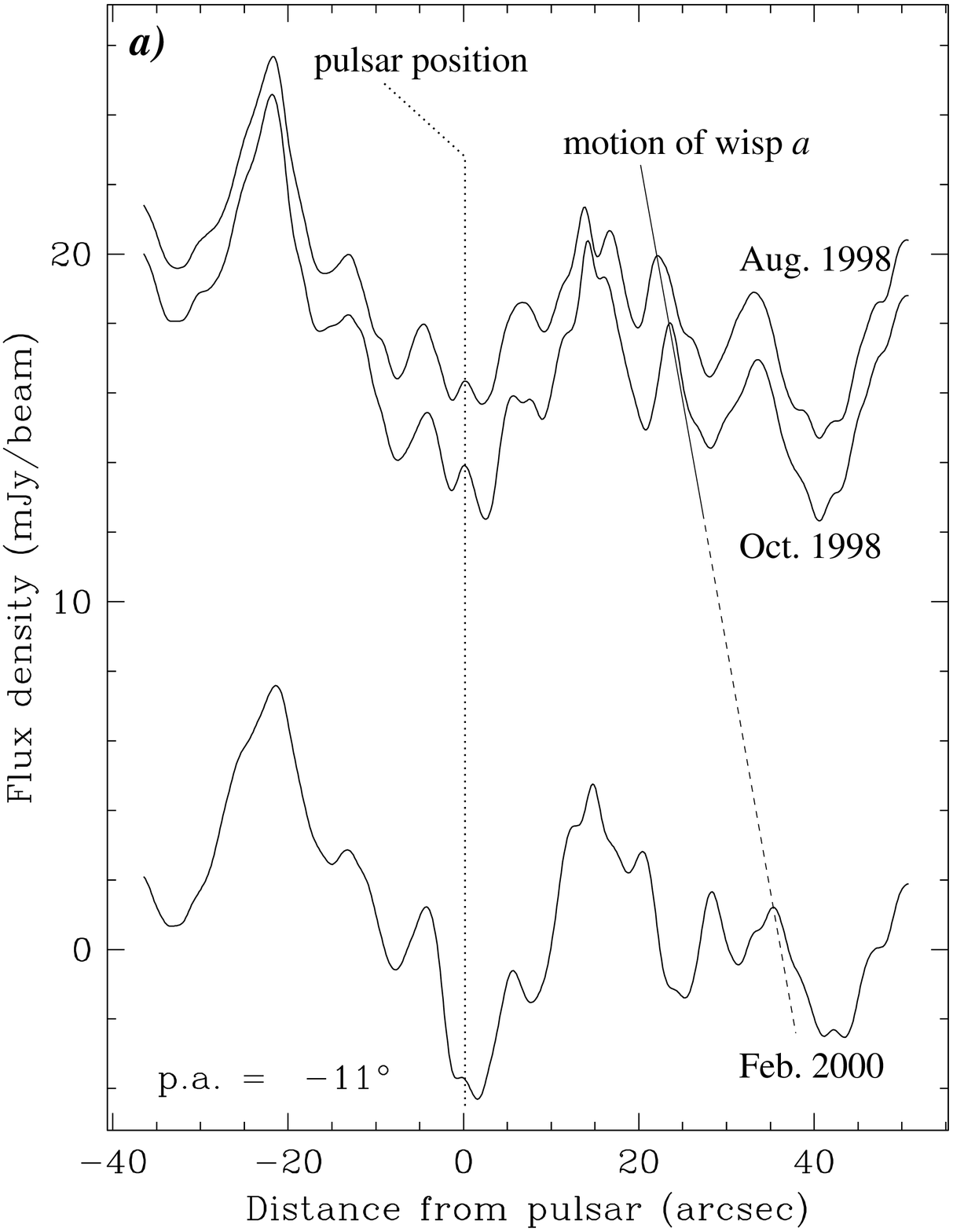}{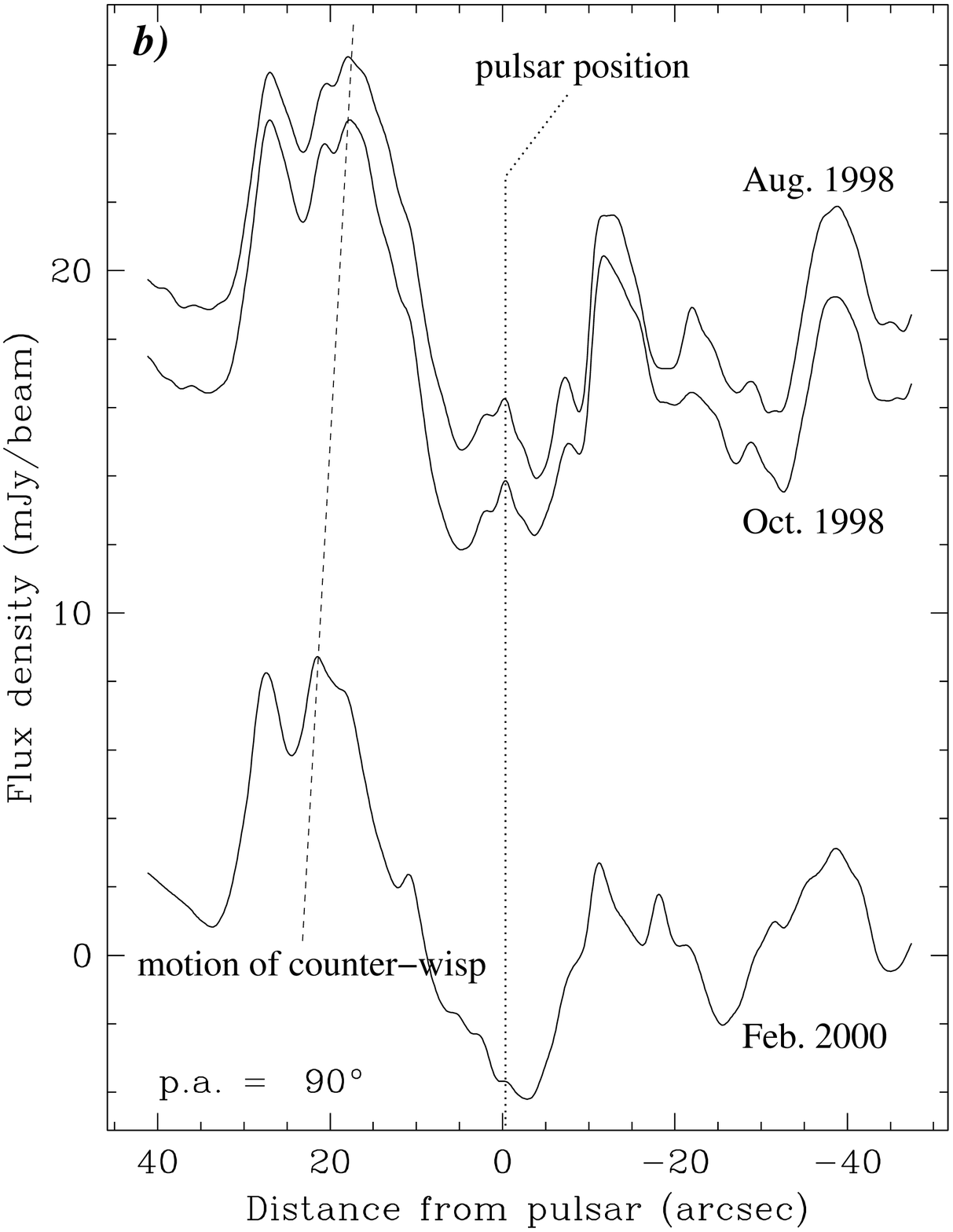}
\figcaption{Profiles through the center of the Crab at three epochs.
The profiles were made from the high-pass filtered images shown in
Fig.~\ref{riplmaps}.  In each case the bottom profile is that
from 2000 February~11.  The middle and the upper profiles are those
from 1998 October~13 and August~9, respectively, and have been
artificially shifted upwards by an amount proportional to the the time
difference to the 2000~February epoch.  The profiles are drawn through
the pulsar position (\Ra{5}{31}{31}{423}, \dec{21}{58}{54}{26}), which
is indicated by the dotted vertical line.  In $a)$ we show profiles
drawn at a p.a.\ of $-11$\arcdeg, with a +ive displacement being to the
NW. The motion of wisp~$a$ is indicated, with the solid line
indicating the secure identification of this feature between 1998
August and October, and the dashed line showing the likeliest
identification in 2000 February.  In $b)$ we show profiles in R.A.\ or
at a p.a.\ of $90$\arcdeg.  The motion of the counter-wisp is
indicated.
\label{profls}}
\end{figure}

Finally, in order measure the speeds of features, we plotted profiles
through the high-pass filtered images.  Figure \ref{profls} shows
profiles for our three epochs, drawn through the pulsar at position
angles, p.a., of $-11$\arcdeg, chosen to pass through the most mobile
wisp, and at $-90$\arcdeg, chosen so as to pass through a mobile
feature on the opposite side of the pulsar.  There are numerous
changes in addition to the motion of wisp~$a$ indicated in
Fig.~\ref{riplmaps} above.  However, with our very coarse sampling in
time, the identification of features from one epoch to the next is
difficult. We note that the pulsar, which is known to be quite
variable but has a mean flux density of 1.4~mJy at 4.8~GHz (Lorimer
et~al.\ 1995), is visible as a variable peak at the origin in the
profiles.

The identification is wisp $a$ between 1998 August and October is
unambiguous.  It moves with an apparent velocity of $0.24 \pm 0.06 c$.
We indicate this motion with a solid line on Fig.~\ref{profls}$a$.  By
2000 February~11, the identification is no longer completely
unambiguous --- we mark the motion derived from the likeliest
identification in Fig.~\ref{profls}$a$ with a dashed line.  This
identification implies an average apparent speed between 1998~August
and 2000~February of $0.28 \pm 0.01 c$, which is consistent with the
speed between 1998 August and October.  On the opposite side of the
pulsar, the most prominent moving feature is seen on the profile at
p.a.\ $= -90\arcdeg$.  We call this feature the counter-wisp, and mark
its motion in Fig.~\ref{profls}$b$.  The speed between 2000 August and
October is not reliably determinable, but the apparent speed between
1998 (taking the average of August and October) and 2000 February is
$0.08 \pm 0.02 c$, and this motion marked in Fig.~\ref{profls}$b$.  We
note that because of the large time interval and the complexity of the
changes between our 1998 and 2000 epochs, the identification of
features over this time-span is necessarily somewhat more tentative.

Assuming an inclination angle of 30\arcdeg, we can calculate the true
speed of the features above: wisp~$a$ has a true speed of $0.38 \pm
0.01 c$, (assuming that the above identification in Feb.~2000 is
correct; see Fig.~\ref{profls}$a$).  The counter-wisp has a true speed
of $0.19 \pm 0.05 c$.

\section{Discussion}

We have shown that there are rapidly moving features, with velocities
$\lesssim c/3$, visible in the radio emission from the Crab Nebula.
These features have an arc-like geometry similar to that of the
optical wisps.  Bietenholz \& Kronberg (1992) had already reported the
existence of similarly shaped features near the center of the nebula,
apparent in the radio spectral index.  They noted the possibility that
these features might in fact be temporal changes rather than true
spectral index features, because the two images from which they
derived the spectral index were effectively taken several months
apart\footnote{The spectral index was between 5~GHz and 1.4~GHz.  The
5~GHz B-array data were taken on 1987 November~25, while the
resolution-matched 1.4~GHz A-array data were taken on 1987 July~7.}.
In light of the present results it seems likely that the features
reported in Bietenholz \& Kronberg (1992) were in fact the result of
temporal changes, rather than true spectral index anomalies.  We
cannot of course exclude the presence of true spectral index
variations, but the data so far at hand do not demand them.
 
We propose that these rapidly variable radio wisps are associated with
the similar features seen in the optical and the torus visible in the
X-ray.  The precise geometry of the radio features is harder to
discern than that of the optical wisps due to the presence of a
relatively much brighter complex background provided by the rest of
the nebula.  Nonetheless our difference images give a distinct picture
of the temporal changes in these features, and clearly show an
elliptical geometry.  In fact, the ellipticity is the same as that of
both the optical wisps and the X-ray torus, suggesting that, like the
optical wisps and the X-ray torus, the radio wisps are circular
features seen at the same inclination angle of $\sim 30\arcdeg$.

Like the optical wisps, the radio wisps seem to be moving outward
(although this is not clearly established by our observations, since
we have only three epochs and sub-optimal time sampling, the
unambiguously identified wisp~$a$ {\em is}\/ moving outwards between
August and October 1998).  Furthermore, we measure speeds of $\lesssim
c/3$ for the features which we can identify from one epoch to the
next.  This is in fact the expected speed for the post-shock flow
(Blandford \& Rees 1974) and consistent with the speeds measured in
the optical (Hester et~al.\ 1996; Hester 1998; Tanvir et~al.\ 1997) and
inferred from the X-ray (Pelling et~al.\ 1987).  These arguments
strongly suggest that the presently observed radio wisps are
associated with the optical wisps and the X-ray torus.

In order to discuss the implications of these observations, we will
briefly describe our theoretical understanding of the Crab Nebula.  As
mentioned in \S~1, the visible synchrotron nebula is powered by the
spindown energy of pulsar.  This energy emerges from the pulsar in the
form of a collimated, highly relativistic wind, which consists of some
combination of particles and magnetic field.  The coupling to the
outer boundary condition, namely the outside of the observable
synchrotron nebula which moves with $v \ll c$, requires a shock in the
wind.  It is at this shock that the particle velocities are
randomized, and it is the post-shock wind which then supplies the
relativistic particles and the magnetic field responsible for the
nebular synchrotron emission. The wisps are thought to be
manifestations of this termination shock, since pressure balance
arguments place the location of the shock roughly at the wisp location
(Rees \& Gunn 1974). Similar features have been seen near other
pulsars (Bietenholz, Frail, \& Hankins 1991; Frail \& Moffett 1993),
and so they may be a common manifestation of the interaction of a
pulsar wind with its surroundings.  In particular, the wisps are
thought to be associated with the shock in the equatorial sector of
the pulsar wind, since both the optical and especially the X-ray
observations (Weisskopf et~al.\ 2000) clearly show a toroidal
geometry. (We note that polar features are also seen in both the
optical and the X-ray, but as yet we cannot conclusively identify any
polar features in the radio).

The precise mechanism of wisp formation, however, is not yet well
understood (Bietenholz \& Kronberg 1992; Chedia et~al.\ 1997; Hester
1998; Arons 1998; Begelman 1999).  Also, while the radio spectral
index is very uniform over the nebula (Bietenholz et al.\ 1997),
suggesting a single and presumably central source for all the radio
emitting electrons, their origin in the shock near the pulsar is not
yet understood (see e.g., Atoyan 1999; Arons 1998).  At the
termination shock of the wind, one expects that the mean Lorentz
factor, $\gamma$, of the post-shock particles to be roughly equal to
that in the wind, which is widely thought to be $\sim 10^6$ (Rees \&
Gunn 1974; Kennel \& Coroniti 1984; Melatos \& Melrose 1996; Arons,
1998).  However, this would produce essentially no radio-emitting,
i.e.\ $\gamma \sim 10^3$ electrons.

For example, the model of Arons (1998; see also Gallant \& Arons 1994)
can account for wisps and also the observed nebular spectrum from the
optical to the X-ray, but it cannot account for the radio-emitting
electrons.  Arons suggests that a slower, high-latitude wind might
produce the radio emitting electrons.  In the model of Atoyan (1999),
which can account for the entire nebular spectrum, the radio emitting
electrons are of historical origin, having been emitted when the
pulsar was younger and spinning much more rapidly.  However, in
neither of these cases would radio-emitting electrons be accelerated
in the current wisp region.  Thus the detection of wisps in the radio
implies that the acceleration of particles in the equatorial shock
termination region produces low energy radio-emitting electrons as
well as those at higher energies responsible for the optical and
higher synchrotron emission.

In summary, we present new radio images of the Crab Nebula, with
higher resolution and dynamic range than those previously available.
We show that there are significant changes occurring over periods of
two months and one year.  An arc-like feature, or wisp, is seen to be
moving outwards, and images of the {\em differences} between
successive pairs of our three observing epochs clearly show numerous
features which have an elliptical geometry.  We propose that these
radio wisps are associated with the optical wisps, which suggests that
acceleration mechanism producing radio-emitting electrons is the same
as, or at least co-spatial to, that producing the more energetic
electrons responsible for the nebular optical and X-ray synchrotron
emission.  Future, more highly time-resolved observations will
hopefully allow us to establish the exact nature and speed of the
motions.

\medskip
\acknowledgements Research at York University was partly supported by
NSERC.  We thank Barry Clark for his patience with the scheduling
difficulties involved in these observations.
  
\onecolumn

\clearpage

%
%

\end{document}